\documentclass[conference]{IEEEtran}
\usepackage[hyphens]{url}
\usepackage{hyperref}
\hypersetup{breaklinks=true,colorlinks,allcolors=blue}
\usepackage[backend=biber, style=ieee]{biblatex}

\usepackage{amsmath,amssymb,amsfonts}
\usepackage{algorithmic}
\usepackage{graphicx}
\usepackage{textcomp}
\usepackage{xcolor}
\def\BibTeX{{\rm B\kern-.05em{\sc i\kern-.025em b}\kern-.08em
    T\kern-.1667em\lower.7ex\hbox{E}\kern-.125emX}}
    
\usepackage{listings}
\usepackage{adjustbox}
\usepackage{caption}
\usepackage{float}
\usepackage{orcidlink}
\lstdefinestyle{myC}{
  language=C,
  basicstyle=\ttfamily\footnotesize,
  stepnumber=1,
  frame=single,
  breaklines=true,
  tabsize=2,
  captionpos=b,  
}

\title{LibLMFuzz: LLM-Augmented Fuzz Target Generation for Black-box Libraries}

\author{\IEEEauthorblockN{Ian Hardgrove\,\orcidlink{0009-0009-7481-5871}}
\IEEEauthorblockA{\textit{The Beacom College of Computer and Cyber Sciences} \\
\textit{Dakota State University }\\
Madison, SD, USA \\
ian.hardgrove@trojans.dsu.edu}

\and

\IEEEauthorblockN{John D. Hastings\,\orcidlink{0000-0003-0871-3622}}
\IEEEauthorblockA{\textit{The Beacom College of Computer and Cyber Sciences} \\
\textit{Dakota State University }\\
Madison, SD, USA \\
john.hastings@dsu.edu}
}

\begin{document}

\maketitle

\floatstyle{plain}
\newfloat{listing}{htbp}{lop}
\floatname{listing}{Listing}

\begin{abstract}
A fundamental problem in cybersecurity and computer science is determining whether a program is free of bugs and vulnerabilities. Fuzzing, a popular approach to discovering vulnerabilities in programs, has several advantages over alternative strategies, although it has investment costs in the form of initial setup and continuous maintenance. The choice of fuzzing is further complicated when only a binary library is available, such as the case of closed-source and proprietary software. In response, we introduce LibLMFuzz, a framework that reduces costs associated with fuzzing closed-source libraries by pairing an agentic Large Language Model (LLM) with a lightweight tool-chain (disassembler/compiler/fuzzer) to autonomously analyze stripped binaries, plan fuzz strategies, generate drivers, and iteratively self-repair build or runtime errors. Tested on four widely-used Linux libraries, LibLMFuzz produced syntactically correct drivers for all 558 fuzz-able API functions, achieving 100\% API coverage with no human intervention. Across the 1601 synthesized drivers, 75.52\% were nominally correct on first execution. The results show that LLM-augmented middleware holds promise in reducing the costs of fuzzing black box components and provides a foundation for future research efforts. Future opportunities exist for research in branch coverage.
\end{abstract}


\section{Introduction}

Over the past several years, fuzzing has become a popular strategy for identifying security vulnerabilities and other reliability issues in binary or native programs \cite{valentin2021}. Fuzzing has a proven track record in discovering vulnerabilities, as demonstrated by the more than 320 vulnerabilities found in open source projects by the OSS-Fuzz program \cite{google/oss-fuzz-vulns_2025}. Fuzzing also avoids the pitfalls of other vulnerability discovery techniques, such as the time investment required for manual analysis and the high false positive rate associated with many static analyzers \cite{6405650}. Even with the positive aspects that have driven adoption, the continual maintenance cost to develop fuzzing targets, and initial investment to achieve coverage on an existing library may deter initial adoption \cite{bohme2020,nourry2025}. 

These trade-offs also impact security researchers attempting to discover and remediate vulnerabilities in applications. The adoption cost of fuzzing is further increased by the fact that security researchers often lack access to the original source code. Estimates suggest that software projects consist of 70-90\% open-source software \cite{A_Summary_of_Census_II}, leaving the remaining up to 30\% as proprietary closed-source legacy code. In order to instrument a fuzzing framework, the researcher needs to develop targets for potentially hundreds of exported functions \cite{Ispoglou_Austin_Mohan_Payer_2020} with a lack of information that is usually present to simplify the development of fuzz targets. 

The popularization of Large Language Models (LLMs), such as through OpenAI's ChatGPT and Google's Gemini, has driven adoption of generative artificial intelligence in a variety of contexts. One such relevant context is the ability of fine-tuned models to generate valid source code within the context of an application. Researchers have found that GitHub's Copilot \cite{github_copilot_2025}, a popular tool in this category, can generate valid source code 90\% of the time \cite{Yetistiren_Ozsoy_Tuzun_2022}. The abilities of LLMs to summarize natural language and generate semantically correct source code lend the opportunity to integrate them into the fuzzing workflow.

In order to address such challenges faced by security researchers, this research looks to use LLMs to generate valid fuzz targets for closed-source binary libraries. Using recent developments in agentic workflows backed by LLMs, the research implements LibLMFuzz, a fuzzing framework that allows an LLM to gather information about a binary library as a means to inform the development of fuzz targets. The capabilities of LibLMFuzz are tested against four commonly used Linux shared libraries to measure the ability to cover exposed Application Programming Interface (API) functions. 
The results demonstrate that LLMs can autonomously synthesis end-to-end fuzz drivers for stripped binaries, providing a foundation upon which next-generation, low-context fuzz-target middleware can be built. 

\section{Background}

\subsection{Fuzzing Libraries}

Where fuzzing was initially developed as a means of testing the reliability of UNIX programs \cite{10.1145/96267.96279}, modern frameworks provide a wide range of features to discover security vulnerabilities. Fuzzing repeatedly calls a function or minimal 
block of a program with random input values in an attempt to find code paths leading to bugs that result in unhandled program states and could cause reliability issues and security vulnerabilities. Tools such as AFL++ \cite{257204} and LLVM's libFuzzer \cite{llvm_libfuzzer_doc} provide custom compiler toolchains, integrations with sanitizers, and the ability to introspect various runtime qualities of the program while fuzzing.

Previous research has shown exhaustive program coverage as a key metric enabling the discovery of software bugs \cite{kochhar2015code}. At a high level, fuzzing has two factors that influence coverage: the quality of randomized inputs to exercise branches conditions during execution, and instrumentation of the program to allow for more unique methods of execution. Generally, fuzzing frameworks measure coverage by using a modified compiler to instrument checkpoints in the compiled program \cite{257204}; however, these features have limited applicability when dealing with pre-compiled binary programs.

A complication specific to fuzzing libraries is that coverage cannot be obtained via a single program entrypoint. Where user-space executable programs generally consist of a single ``main" function acting as the entrypoint, libraries are instead composed of one or more exported functions acting as isolated entrypoints. Reaching high percentages of coverage therefore requires that each entrypoint has a unique target generated to execute its functionality.

\subsection{Large Language Models and Agents}

Current LLMs are a specialization of the original generative pre-trained transformers (GPTs) \cite{Radford_Narasimhan_Salimans_Sutskever_2018}, which themselves are a specialization of transformer-based neural networks \cite{Devlin_Chang_Lee_Toutanova_2019}. Such neural networks are comprised of complicated architectures and often utilize billions of parameters based on their training corpus, target size, and performance metrics \cite{Introducing_Llama_3_1}. Each prompt provided to the model is processed by this multi-layer system as a means of determining the most likely next token based on its training data.

The ability to process natural language has resulted in applications outside of conversational AI. Applications such as the previously mentioned Copilot enable modification and extension of existing software projects via code completion and autonomous code generation. These tools extend beyond historical static analysis techniques as the models have access to the entire source tree to use as context when autonomously generating code \cite{githubdocs}. Code generation is highly relevant as fuzz targets are intrinsically specialized programs requiring semantic understanding to call into existing code.

An exciting advancement in the use of LLMs has been the ability to integrate with third-party tooling, giving the models agency to perform actions on a user's behalf. The ReAct agent architecture has shown significant improvement over Chain-of-Thought and other model interaction methods by allowing the model to retrieve data specific to the user's prompt \cite{yao2022react}. Implementations of agents use structured messages between a LLM and framework to execute developer-implemented functionality. One such example is Google's Gemini and it's ability to utilize Google Search to gather additional information related to a prompt.

\section{LibLMFuzz} 

LibLMFuzz implements a middleware providing an integration point between the end-user, LLM, and the functions providing additional context to the model. This middleware manages multiple phases of execution, prompting the model based on the phase to suggest useful tools and provide coverage over the entirety of the exported API. An implicit phase 0 provides the middleware with the basic information allowing coverage over the shared object's exported functions. Phase 1 begins iterative interaction with the LLM to gather information as input for fuzz target generation. Finally, Phase 2 performs code generation of targets, error correction, and validation of model generated fuzz target.

\begin{figure}[h]
    \centering
    \includegraphics[width=1.0\linewidth]{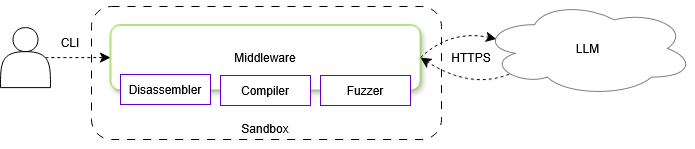}
    \caption{Middleware Infrastructure Architecture}
    \label{fig:infrastructre}
\end{figure}

As tools made available to the model by the middleware allow for execution of selective software programs on the host, the middleware executes within a sandboxed environment visually represented in Fig. \ref{fig:infrastructre}. The series of tools, a disassembler, compiler, and fuzzing engine, are each installed in the same virtualized Linux host as the middleware. Although underlying tools are independent and modular, radare2 provides disassembly, and LLVM's clang and libFuzzer provides the compiler and fuzzing engine, respectively.

The LangChain library \cite{langchain_0_3_19_2025} is used to provide web API access to models and simplify integration with tools. Although the middleware's modularity for the model API allows locally hosted models to be used for development, Google's Gemini 2.0 Flash was used for finalizing development and results gathering.

\subsection{Implementation Workflow}

\begin{figure}[h]
    \centering
    \includegraphics[width=1.0\linewidth]{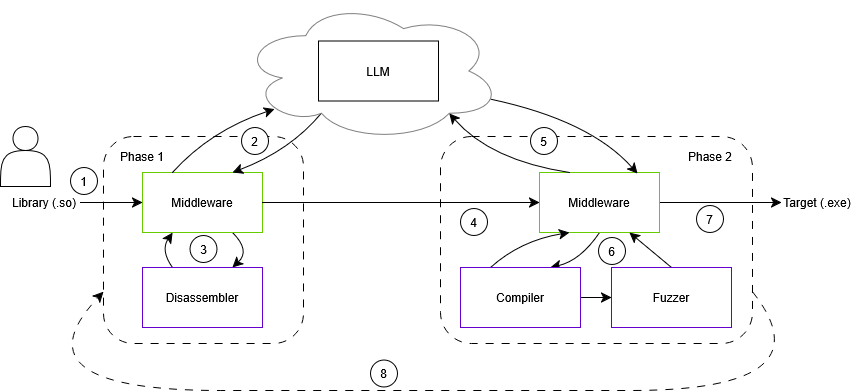}
    \caption{Workflow of Library Fuzz Target Generation}
    \label{fig:workflow}
\end{figure}

LibLMFuzz provides a general framework by using multiple prompts to separate the information gathering and target generation into distinct phases of execution. Within each phase, the middleware gives the model autonomy to interact with the available tools to gather information about the fuzz target. Fig. \ref{fig:workflow} provides a visual representation of LibLMFuzz's workflow and integration points.

\begin{enumerate}
    \item The user executes LibLMFuzz by providing it with a binary shared object for which fuzzing targets should be generated. The middleware then invokes the disassembler to generate a list of exported functions, filtering the list for fuzz-able functions which take user input. 
    \item The middleware then starts communication with the remote LLM, prompting it to gather information about an individual exported function.
    \item Based on the model's response, the middleware then may reach out to the disassembler to gather the functions disassembly and the derived function signature. This also creates the first feedback loop, as the model has the ability to repeat steps 2 and 3 as necessary.
    \item Phase 2 transitions the priority of the middleware to having the model generate a fuzzing target. The model can no longer return to the analysis loop of steps 2 and 3 for the given function after this point.
    \item The model is re-prompted with a new prompt focusing on generation of a fuzz target based on the previously gathered context.
    \item Based on the model response, the middleware attempts to write any generated C/C++ code for the fuzzing target to the file system and compile the target. Once successfully compiled, the fuzz target is executed for a short period of time to ensure it is both syntactically and nominally valid. If either compilation or initial execution fail, the model is re-prompted with the error information to fix any mistakes. 
    \item Once all prior steps have successfully completed, the middleware saves the information and exits the prompting loop with the model.
    \item If any exported function has not had a fuzz target generated, the middleware returns to Phase 1 with the next function as input for the prompt.
\end{enumerate}

In communicating with the LLM, several prompt types and strategies are implemented to receive useful responses. A system prompt provides continuous high-level context to the model, asking it to act as a security researcher with a focus on fuzzing. Prompts used at individual steps in the process attempt to focus the model on the current task, avoid commonly discovered pitfalls, and receive usable formatted output where necessary. For example, when generating the source code for a fuzz target, the prompt provides information regarding the command that will be used to compile the target, specifies that the model avoid data structures not found in the binary analysis, and requests that only source code is output.

\subsection{Error Correcting Feedback Loops}

LibLMFuzz incorporates multiple feedback loops to provide opportunities for the model to correct erroneous output. These cycles were  included to validate the syntax and semantics of model generated fuzz targets. For example, the two-step phase of both compiling and executing compiled targets for a ten second span provides feedback about the validity of the syntax and the correctness of the semantics of the target. If compilation failed for any reason, the model receives the literal compiler error, enabling it to autonomously correct the errors. The short execution provided similar feedback, allowing the model to revise the code in cases where early failures could be remediated to improve the quality of the fuzz target.

\section{Methodology}

To empirically study the capabilities of LibLMFuzz, we devised a test methodology to determine the success rate of target generation. LibLMFuzz was evaluated on four widely used open-source libraries (see Table \ref{table:results}), selected because they are popular benchmarks for effective fuzz testing. 
In order to match the goals of the project while providing some real-world variability (i.e., compiler flags, optimization levels, inclusion of symbols), each open-source library was compiled into its binary, shared object form. LibLMFuzz was then run against each of the compiled shared objects to generate targets for the discovered fuzz-able functions. 

The framework was installed within a containerized environment due to its ability to execute tools directly on the host. This container environment was allocated two cores on an x64 system, 2 GB of RAM, with internet access provided to communicate with the LLM. 
Due to the varying number of exported function and library complexities, no overall time limit was enforced on interaction with Google's Gemini Flash 2.0. Additionally, the vast majority of executions, including larger, more complex libraries, could be completed within a single day of the free-tier rate limiting enforced by Google.

\section{Results \& Discussion}

Table \ref{table:results} shows the API coverage statistics for each of the open-source libraries. With a total of 558 exported functions deemed fuzz-able by the middleware, LibLMFuzz was impressively able to generate a syntactically valid target for 100\% of the cases. Of the 1601 total generated source code targets, 75.52\% were nominally valid and passed the short execution test. This equates to 2.87 source code targets per fuzzable function in order to reach an executable target.

\begin{table}[h!]
    \caption{Results of Running Shared Objects in LibLMFuzz}
    \label{table:results}
    \centering
    \begin{tabular}{|c c c c c|} 
        \hline
                & Fuzzable & Target & Compiled & API \\
        Library & Exports & Source Code & Targets & Coverage \% \\ [0.5ex] 
        \hline 
        cJSON & 144 & 274 & 170 & 100 \\ 
        libmagic & 32 & 104 & 96 & 100 \\
        libpcap & 200 & 602 & 405 & 100 \\
        libplist & 182 & 621 & 538 & 100 \\
        \hline
        Total & 558 & 1601 & 1209 & 100 \\
        \hline
    \end{tabular}
\end{table}

While LibLMFuzz showed an overall high success rate at generating API coverage, the quality of the targets generated by the middleware lacked the understanding of program semantics to reach ideal branch coverage. Early design decisions made the ability to measure these pitfalls immeasurable; modifications to a fuzz engine would be required to gather the additional statistics. The following section aims to address how the design lead to both high API coverage but low semantic program understanding.

\subsection{Lack of Requisite Context}

A significant challenge when prompting an LLM to generate a fuzz target for a given function is providing enough semantic information about the library. Although the results showed a $\sim$75\% success rate in terms of compilable targets also being semantically valid, this result is tempered by missing statistical information on program branch coverage from the generated targets. In order to gather initial data, a disassembler was integrated to allow the model to gather function definitions, including the assumed type information, and the disassembly for the function.  In nearly all cases, the type information for the function signature was incomplete, such as libraries where the build toolchain stripped debug type information, therefore functions using pointers were often represented as 64-bit integers due to the AMD64 architecture used by the sandbox environment.

The model subsequently struggled to derive additional type information from the assembly unless common functions such as ``memcpy" or string-related operations were visible. This led to many fuzz targets for functions accepting complex types by reference being degraded to fuzzing only the address of the pointer. Listing \ref{list:poor-quality-example} is one example where the limited usage context allowed the model to generate a syntactically valid target, but missed the legitimate semantics of the API. The actual signature is added as a comment before the legitimate signature used by the LLM. The difference in type information results in a drastically limited diversity of inputs during the fuzzing process.

\begin{figure}[h!]
\captionsetup{type=listing}
\begin{adjustbox}{width=1.3\columnwidth, margin={0.3cm 0cm 0cm 0cm}}
\begin{lstlisting}[style=myC]
#include <cstdlib>
#include <cstdio>
#include <cstdint>
#include <cstddef>
#include <cstring>

// CJSON_PUBLIC(char *) cJSON_Print(const cJSON *item)
extern "C" int64_t cJSON_Print(int64_t arg1);

extern "C" int LLVMFuzzerTestOneInput(const uint8_t *Data, size_t Size) {
    if (Size < 8) {
        return 0;
    }
    int64_t arg1 = 0;
    std::memcpy(&arg1, Data, 8);
    cJSON_Print(arg1);
    return 0;  // Values other than 0 and -1 are reserved for future use.
}   
\end{lstlisting}
\end{adjustbox}
\captionof{listing}{Example of poor disassembly context leading to a low quality fuzz target.}
\label{list:poor-quality-example}
\end{figure}

An additional limitation was the inability to provide the model with context for the library as a whole. Whereas many functions in the public facing API may emulate object-oriented access to transient structures, the model was only formally made aware of a single function during each fuzz cycle. Therefore, even if the model were able to infer that a function parameter was a pointer, it lacked context about other functions that may be responsible for creating an object, even when the structure’s definition was available. 

\subsection{Hallucinations via Leak of Training Data}

An interesting side-effect based on experimental configuration was the hallucination and leaking of training data.  One downside of using compiled versions of popular open-source libraries was that the model intrinsically contained information about the usage patterns and type information. Early in development and prompt engineering, a fuzz target would routinely fail compilation due to header files and structure types and definitions being added by the model. The headers and data structure were often valid in normal usage, but were not made available at compile time to represent the real-world challenges of working with closed-source binary libraries.

Additional considerations were integrated into prompts to mitigate the effect of hallucinations represented as training data leaks. Statements in the prompt directly asked that the model not assume information about the function signature. Subjectively, the downstream effects of this were two-fold: the model ceased injecting information available via its training data, but also became more conservative about deriving type information. This resulted in a higher success rate when compiling, but likely reduced the potential quality of model output as in the form of derived type information in the target.

Listing \ref{list:example-hallucinated-functions} shows an example of the LLM being able to add context not provided by the middleware or disassembler. When given the function \verb|cJSON_ParseWithLengthOpts| as an exported function for which to make a target, the model inserts another legitimate function in the extern section, showing additional knowledge of API usage.

\begin{figure}[h!]
\captionsetup{type=listing}
\begin{adjustbox}{width=1.3\columnwidth, margin={0.3cm 0cm 0cm 0cm}}
\begin{lstlisting}[style=myC]
#include <cstdlib>
#include <cstdio>
#include <cstdint>
#include <cstddef>

extern "C" {
    int64_t cJSON_ParseWithLengthOpts(const char *arg1, size_t arg2,
        const char **arg3, int arg4);
    void cJSON_Delete(int64_t item);
}

extern "C" int LLVMFuzzerTestOneInput(const uint8_t *Data, size_t Size) {
    const char *arg1 = (const char *)Data;
    size_t arg2 = (size_t)Size;
    const char *arg3_ptr;
    const char **arg3 = &arg3_ptr;
    int arg4 = 0;
    
    int64_t json = cJSON_ParseWithLengthOpts(arg1, arg2, arg3, arg4);
    if (json) {
        cJSON_Delete(json);
    }
    return 0;
}    
\end{lstlisting}
\end{adjustbox}
\captionof{listing}{Example of hallucinated functions likely being pulled from training data.}
\label{list:example-hallucinated-functions}
\end{figure}

\subsection{Prompt Engineering}

As briefly mentioned in earlier sections, continuous prompt refinement was critical in achieving API coverage. Developing a successful prompt was an iterative process requiring continual improvement in order to have the model successfully perform the desired actions. This also varied on a per-model basis, as locally hosted models used for initial development responded differently to prompts than Gemini 2.0 Flash. Including commands in the prompt that directed the model to derive but not guess at types helped solve issues with hallucinations, and additional directions to omit security mitigations kept fuzz target source code simple and compatible with a fuzzing engine.

One attempted strategy early in development was the use of a single planning prompt. Planning describes the set of desired actions as part of the initial prompt and allows the model to perform each action using the order and method described. Using this method of prompting with both the locally hosted and Gemini 2.0 Flash models regularly resulted in failed attempts to generate targets, and in skipping required steps to gather initial context prior to attempting to generate the target. Using a phased approach to prompting the model, especially as part of tool usage, resulted in significant performance and quality benefits.

\section{Related Work}

Prior research on automated fuzzing appears in two main categories. The first category generates fuzz drivers through static or dynamic analysis of source-available code. The second category utilizes LLMs to boost either input generation or driver construction, also relying on access to the program’s source. Notably, the literature contains no end-to-end approach that generates fuzz targets for binary-only, closed-source libraries, the gap that LibLMFuzz addresses. 

\subsection{Automated Fuzz Target Generation}

Autonomous generation of fuzz targets precedes the popularity of integrating one or more LLMs into the generation process. Research in this category generally falls into one of two categories: the application of static analysis as a basis for target generation, or runtime introspection to gather conventional usage and type information. Several approaches have shown success using domain-specific languages with knowledge of static analysis properties of source-available programs to generate fuzz targets \cite{Ispoglou_Austin_Mohan_Payer_2020,Chen_Xie_Lyu_Wang_Chen_2023}, with one project achieving a 99.8\% target generation success rate and branch coverage approaching a 50\% average across the sampled projects. Alternative runtime-based strategies \cite{Zhang_Lin_Li_Xue_Xie_Chen_Ying_Wang_Liu_2021} showed that dynamic tracing provided sufficient context to successfully generate targets for SDKs with complex function dependency conventions.

\subsection{Augmenting Fuzzing with LLMs}

Researchers have developed numerous strategies to integrate LLMs into both the input and target generation processes of fuzzing. Various research \cite{298230,10546625} has shown success integrating LLMs into the input generation strategy which allowed for fuzzing complex software, including emulation of a CPU's instruction set architecture to find vulnerabilities in the implementation. \cite{Xia_Paltenghi_Le_Tian_Pradel_Zhang_2024} presented novel integration of multiple symbiotic LLMs, allowing fine-tuned models to increase the relevancy of inputs to the software under test.

Researchers have iteratively improved strategies allowing LLMs to develop fuzz targets for open-source projects. Initial work in this category, performed by Google's Open Source Security team, developed the strategy of an autonomous middleware receiving introspected coverage statistics to increase the coverage of existing projects in their OSS-Fuzz project \cite{Fuzz_target_generation_using_LLMs}. The extended coverage enabled rediscovery of recent vulnerabilities in the OpenSSL project that prior targets were unable to discover.  PromptFuzz built upon Google's concept and improved branch coverage by 1.6\% by using valid function executions to enhance prompt context \cite{Lyu_Xie_Chen_Chen_2024}. While these strategies inspired the work performed in LibLMFuzz, prior work focused on source-available projects and does not account for binary and closed-source programs.

\section{Future Work}

The approach introduced by LibLMFuzz provides several opportunities for future developments. One possible method to improve contextual awareness of the model is to combine the existing agentic workflow with a Human-in-the-Loop (HITL) strategy. In terms of improving the quality of the context, the middleware could ingest information from a security researcher who is reverse engineering the application. Additional details provided by reverse engineering, such as function signature type information and function and variable names would provide additional input data that could improve the semantics of the model generated targets.

Similarly, integration of additional tools, providing the LLM with further agency to gather more context through additional channels would present possible opportunities to improve target quality. Giving the model additional tools to execute and introspect function using dynamic debugging, or simply further explore the library contents beyond the function under test may naturally enable better conventional API usage.

An ablation study of the prompts used by the middleware would provide further insight into the value of the prompts in generating valid targets and avoiding hallucination. The prompt used to gather the results in this paper was refined numerous times throughout development, as early iterations had significantly higher failure rates. An overall ablation study considering the prompts and important context would help to define which target development areas would have the greatest impact on success.

\section{Conclusion}

This research presented LibLMFuzz an early approach to applying LLMs to the task of developing fuzz targets for closed-source libraries. The developed middleware provides an integration point, allowing an LLM to gather information and use feedback loops to autonomous generate syntactically and semantically valid fuzz targets. The success in reaching full coverage over the public API of all tested libraries shows potential, with opportunities to improve in context derivation, and additional research into how various forms of context impact the quality of generated targets.

\printbibliography

\end{document}